\documentstyle[12pt,epsf]{article}
\begin{document}
\title{ \bf A simplified
model for monopole catalysis of nucleon decay}
\author{
{\large Y. Brihaye}$^{\diamond}$, 
{\large V. A. Rubakov}$^{\ddagger}$, 
{\large D. H. Tchrakian}$^{\star}$
and {\large F. Zimmerschied}$^{\dagger}$ \\ \\
$^{\diamond}${\small Physique-Math\'ematique, Universit\'e de 
Mons-Hainaut, Mons, Belgium}\\ \\
$^{\ddagger}${\small Institute for Nuclear Research of the
Russian Academy of Sciences,} \\
{\small 60th October Anniversary Prospect, 7a, 117 312, Moscow,
Russia}\\ \\
$^{\star}${\small Mathematical Physics, National University of Ireland
Maynooth (NUIM)} \\
{\small Maynooth, Co. Kildare, Ireland}
\\ \\
$^{\dagger}${\small Fachbereich Physik, Universit\"at Kaiserslautern,
 Postfach 30 49} \\
{\small Erwin-Schr\"odinger-Strasse, Bau 46, D-67663 Kaiserslautern,
Germany}}

\date{}
\newcommand{\dd}{\mbox{d}}\newcommand{\tr}{\mbox{tr}}
\newcommand{\ee}{\end{equation}}
\newcommand{\eea}{\end{eqnarray}}
\newcommand{\be}{\begin{equation}}
\newcommand{\bea}{\begin{eqnarray}}
\newcommand{\ii}{\mbox{i}}\newcommand{\e}{\mbox{e}}
\newcommand{\pa}{\partial}\newcommand{\Om}{\Omega}
\newcommand{\vep}{\varepsilon}
\newcommand{\bfph}{{\bf \phi}}
\newcommand{\lm}{\lambda}
\def\theequation{\arabic{equation}}
\renewcommand{\thefootnote}{\fnsymbol{footnote}}
\newcommand{\re}[1]{(\ref{#1})}
\newcommand{\R}{{\rm I \hspace{-0.52ex} R}}
\newcommand{\N}{{\sf N\hspace*{-1.0ex}\rule{0.15ex}%
{1.3ex}\hspace*{1.0ex}}}
\newcommand{\Q}{{\sf Q\hspace*{-1.1ex}\rule{0.15ex}%
{1.5ex}\hspace*{1.1ex}}}
\newcommand{\C}{{\sf C\hspace*{-0.9ex}\rule{0.15ex}%
{1.3ex}\hspace*{0.9ex}}}
\newcommand{\eins}{1\hspace{-0.56ex}{\rm I}}
\renewcommand{\thefootnote}{\arabic{footnote}}

\maketitle
\begin{abstract}
We present a simple model where a nucleon is 
treated as the Skyrmion,  both monopole 
and Skyrmion are smooth solutions of the classical field equations,
the baryon number non-conservation is explicitly built in and is
due to the classical analogue 
of the triangle anomaly,
and there is no necessity to consider leptons. 
We show by numerical analysis that there are no
static classical solutions with non-zero baryon number
on top of the monopole, i.e., that the monopole
catalysis of Skyrmion decay indeed proceeds classically.
\end{abstract}
\medskip
\medskip
\newpage

\section{Introduction}

An interesting feature of magnetic monopoles in gauge theories
is that they can catalyse nucleon decay at strong interaction rate
\cite{rubakov,callan}. The actual computation of the monopole
catalysis cross section remains, however, an open problem.
This computation would involve long-distance physics of two 
different kinds --- quark-monopole interactions and QCD phenomena ---
and also short-distance physics responsible for baryon number
non-conservation. Most of the relevant effects are non-perturbative,
and this makes even a semi-qualitative  calculation difficult.

An approach to this problem was suggested in 
Ref.~\cite{callanwitten}. It was proposed to invoke the Skyrme 
model of a nucleon \cite{skyrme} which indeed works reasonably 
well in describing low energy properties of baryons 
\cite{adkins,skyrmionreview}. In the Skyrme model,  the nucleon
is essentially
a classical field theory object, a soliton. Given that the 
't~Hooft--Polyakov monopole is also a soliton, the 
monopole-nucleon interactions are then possible to describe 
at the level of classical field theory.

In the proposal of Ref.~\cite{callanwitten}, the monopole is 
treated as point-like. The non-conservation of baryon number 
comes in through the boundary condition imposed on
the Skyrme field at the monopole centre, i.e., 
at the singularity.
It is not entirely obvious whether concrete mechanisms of
baryon number non-conservation indeed produce,  
in the limit of vanishing monopole size, the conjectured
boundary condition.
Also, dealing with the singularity and the boundary condition 
imposed at the singularity may be inconvenient for 
numerical computations.  One more complication inherent in
the proposal of   Ref.~\cite{callanwitten} stems from
the fact that the spherically symmetric Skyrme
field in the monopole background corresponds to an electrically
charged object (``proton''), so one has to explicitly
introduce leptons to make baryon number violation
consistent with electric charge conservation.

In this paper we present a simplified model
where both monopole 
and Skyrmion are smooth solutions of the classical field equations,
the baryon number non-conservation is explicitly built in and is
due to the classical analogue \cite{wittenwz,dhoker} 
of the triangle anomaly,
and there is no necessity to consider leptons.
The price to pay is that the Skyrme field in the presence of
the monopole does not carry net electric charge, so our model
mimicks the monopole catalysed decay of a neutron, rather than 
a proton.

We shall concentrate on two aspects of our model. First, 
we study whether or not the baryon number non-conservation
proceeds without suppression at the classical level. 
For this we shall have to consider a region near the
monopole core. Of course, it is not realistic to use the
non-linear sigma model for describing physics near the
core, but this step is inevitable in the classical
calculation of monopole catalysis. By carrying out numerical analysis,
we show that there are no states with non-zero baryon number on top
of the 't~Hooft--Polyakov monopole, irrespectively of the size of the
monopole core, i.e., monopole catalysis of nucleon decay
indeed takes place. 

Second, we find that the regime of ``point-like'' 
monopole sets in when the two scales, responsible for the
sizes and masses of the monopole and Skyrmion, respectively,
are not yet widely different:
our results are stable when the ratio of monopole and Skyrmion
sizes is below 0.03, and the mass ratio is below 1.
This is a good sign for future numerical analysis of the
Skyrmion decay in Skyrmion-monopole collisions. Hence, our model
may be a reasonable starting point for a semi-realistic 
calculation of the monopole catalysis cross section.

This paper is organized as follows. We present our model in
Section 2. To set the stage, we recapitulate the properties of
the monopole and the Skyrmion in Section 3. Section 4 contains our
main results, both analytical and numerical. We present
concluding remarks in Section 5.

\section{The model}

To motivate our model, we recall first that the simplest
theory incorporating Skyrmions has $SU(2)_L \times SU(2)_R$
global symmetry and contains the sigma field $U(x)$ 
taking values in $SU(2)$. We wish to introduce baryon number
non-conservation through the classical anomaly in the baryonic 
current. The baryon number will be anomalous if $SU(2)_L$ or 
$SU(2)_R$ or both are gauged. The model should also possess 
magnetic monopoles; if we do not extend the symmetry group, 
the monopoles are to be associated with the gauged (part of)
$SU(2)_L \times SU(2)_R$. Gauging $SU(2)_L$ alone is not sufficient:
in that case the Skyrme field would break the gauge group 
completely, no electromagnetic $U(1)$ would be left unbroken,
and there would be no magnetic monopoles. So, we come to 
fully gauged $SU(2)_L \times SU(2)_R$ with gauge potentials
\[
    SU(2)_L :  \quad A_{\mu}
\]
\[
     SU(2)_R : \quad B_{\mu}
\]
Realistic monopoles have very small sizes, so we have to break
the non-Abelian symmetry down to an Abelian subgroup at a 
high energy scale. Thus, we introduce two Higgs triplets
$\Phi_A$ and $\Phi_B$ in $({\bf 3}, {\bf 1})$ and 
$({\bf 1}, {\bf 3})$ representations, respectively. These break
$SU(2)_L \times SU(2)_R$ down to $U(1)_L \times U(1)_R$. The
electromagnetic group is the vectorial subgroup of this
$U(1)_L \times U(1)_R$, so we have to break axial $U(1)$ at
some point. For simplicity, we consider the minimal option, 
i.e., we do not introduce extra Higgs fields. Then the axial
$U(1)$
is broken by the Skyrme field itself. The existence of 
a relatively long-ranged axial gauge field
is an unrealistic feature of our
model; however, 
in this paper we study the dynamics near the monopole core,
so it will not be important whether or not the
axial $U(1)$ is truly long ranged. It is straightforward 
to extend our model by introducing yet another Higgs field
which would break the axial $U(1)$, but not the
electromagnetic subgroup.

To summarise, our model has $SU(2)_L \times SU(2)_R$ gauge 
symmetry, with the Skyrme field and two Higgs fields in
the following representations
\begin{eqnarray}
  U \in SU(2)\; , \;\;\; U&:& \;\;({\bf 2}, {\bf \bar{2}}) \nonumber \\
    \Phi_A &:& \;\; ({\bf 3}, {\bf 1}) \nonumber \\
    \Phi_B &:& \;\; ({\bf 1}, {\bf 3})
\end{eqnarray}
To simplify the calculations, we take the gauge couplings, 
Higgs self-couplings and vacuum 
expectation values of the Higgs fields the same in the left
and right sectors,
\[
    g_A = g_B \equiv g \; , \;\;\;
    \lambda_A = \lambda_B \equiv \lambda \; , \;\;\;
    v_A = v_B \equiv v
\]
The action for this model is
\begin{eqnarray}
     S &=& \int~d^4x~\left[ -\frac{1}{2g^2} \mbox{Tr}(F_{\mu\nu}^2)
 -\frac{1}{2g^2} \mbox{Tr}(G_{\mu\nu}^2)\right] \nonumber \\
  &-& \int~d^4x~\left[
\frac{1}{2} \mbox{Tr}(D_{\mu}\Phi_A)^2 + 
\frac{1}{2} \mbox{Tr}(D_{\mu}\Phi_B)^2
\right]
\nonumber \\
&-&  \int~d^4x~\left[
\frac{\lambda}{8} 
\left(\frac{1}{2} \mbox{Tr}(\Phi_A^2) + v^2\right)^2
+  \frac{{\lambda}}{8} 
\left(\frac{1}{2}\mbox{Tr}(\Phi_B^2) + v^2\right)^2
\right] \nonumber \\
&+& \int~d^4x~\left[  
- \frac{F_{\pi}^2}{16} \mbox{Tr}(U^{\dagger} D_{\mu} U)^2
 + 
\frac{1}{32e^2}\mbox{Tr}([U^{\dagger}D_{\mu}U,U^{\dagger}D_{\nu}U]^2)
\right] \nonumber \\
&+& \Gamma_{WZW}
% + \int~d^4x~ \epsilon^{\mu\nu\lambda\rho} Z^{\mu\nu\lambda\rho} 
\label{action}
\end{eqnarray}
where $F_{\mu\nu}$ and $G_{\mu\nu}$ are the field strengths
of $A_{\mu}$ and $B_{\mu}$, the fields $\Phi_A$ and $\Phi_B$
are $2 \times 2$ matrices from the algebras of $SU(2)_L$ and 
$SU(2)_R$, respectively,
\begin{eqnarray}
    D_{\mu}U &=& \partial_{\mu} U + A_{\mu}U - UB_{\mu}
      \nonumber \\
    D_{\mu} \Phi_A &=& 
     \partial_{\mu}\Phi_A + [A_{\mu}, \Phi_A]
\nonumber \\
D_{\mu} \Phi_B &=& 
     \partial_{\mu}\Phi_B + [B_{\mu}, \Phi_B]
\end{eqnarray}
and $F_{\pi}$ and $e$ are the pion decay constant and the Skyrme
constant, respectively. The last term in eq.~(\ref{action}) 
is introduced to reflect anomalies \cite{wittenwz}. Unlike in
Ref.~\cite{callanwitten}, this term has almost no effect in our 
model. In most of this paper we neglect this term, and discuss 
its effect towards the end.

In complete analogy to Refs.~\cite{wittenwz,dhoker}, the 
gauge-invariant baryonic current is
\begin{eqnarray}
   j_\mu &=&
\frac{1}{24\pi^2}\varepsilon_{\mu \nu \rho \sigma}
\mbox{Tr}(U^{\dagger}D_{\nu} U D_{\rho} U^{\dagger}D_{\sigma} U)
\nonumber \\
& +& \frac{1}{16\pi^2}\varepsilon_{\mu \nu \rho \sigma}\mbox{Tr}
(F_{\rho \sigma}D_{\nu}UU^{\dagger}+G_{\rho \sigma}U^{\dagger}D_{\nu}U)
\label{8}
\end{eqnarray}
In the absence of the gauge fields, $A_{\mu}=B_{\mu} =0$,
this current reduces to the topological current,
\[
   k_{\mu} = \frac{1}{24\pi^2}\varepsilon_{\mu \nu \rho \sigma}
\mbox{Tr}(U^{\dagger}\partial_{\nu} U \partial_{\rho} 
U^{\dagger}\partial_{\sigma} U)
\]
while in the presence of gauge fields it has an anomaly,
\[
    \partial_{\mu}j_{\mu}=
\frac{1}{16\pi^2}\mbox{Tr}(\tilde F_{\mu \nu}F_{\mu \nu}
-\tilde G_{\mu \nu}G_{\mu \nu}) 
\]
Thus, topologically non-trivial gauge fields lead to 
the non-conservation of baryon number; in particular, they can 
unwind the Skyrmion (cf. Refs.~\cite{dhoker,Rubakov:1985ba,ambjorn}).

\section{Monopole and Skyrmion}
If it were not for the Skyrme field $U$, the left and right
sectors of this model would decouple, the model would reduce to
two identical copies of the Georgi--Glashow $SU(2)$ model and
there would exist two types of 't~Hooft--Polyakov monopoles.
In the presence of the Skyrme field, the axial $U(1)$ is broken,
and there exists only one type of monopole. The monopole 
solution with zero baryon number has $U({\bf x}) =1$ everywhere in
space, whereas $A_i({\bf x})=B_i({\bf x})$, 
$\Phi_A({\bf x})=\Phi_B({\bf x})$ are precisely the monopole 
fields of the Georgi--Glashow model. The monopole size is
determined by the mass of the vector bosons; at $v \gg F_{\pi}$
one has $m_V = 2gv$, so that
\begin{equation}
     r_{mon} = \frac{1}{2gv}
\label{12*}
\end{equation}
At large $v$, the monopole mass is twice the mass
of the $SU(2)$ monopole, i.e.,
\begin{equation}
   M_{mon} = 4\pi D_{mon} \frac{v}{g}
\label{12**}
\end{equation}
where the numerical constant $D_{mon}$ depends slightly on
${\lambda}/g^2$ and at ${\lambda}/g^2 = 0.5$ is equal to
\cite{marinov}
\begin{equation}
  D_{mon}({\lambda}/g^2 = 0.5) \approx 2.4
\label{12+}
\end{equation}
The estimates (\ref{12*}), and (\ref{12**}) and (\ref{12+}) will
serve as reference points for our study of the monopole--Skyrmion
system.

In the regime of weak gauge coupling, the model possesses the
Skyrmion. Once the gauge fields are ignored, this is the
standard Skyrmion of the $SU(2)$ sigma model. Its radius, as
defined in Ref.~\cite{adkins}, 
is
\begin{equation}
    r_{Sk} = \frac{2.1}{eF_{\pi}}
\label{rSk}
\end{equation}
and its mass is equal to
\begin{equation}
      M_{Sk} = 4 \pi D_{Sk} \cdot \frac{F_{\pi}}{e} \; ,
\;\;\; D_{Sk} \approx 2.9
\label{VV}
\end{equation}
To mimic the situation existing in Grand Unified Theories,
in what follows we shall mostly consider the range of 
parameters in which the ratios $r_{mon}/r_{Sk}$ and $M_{Sk}/M_{mon}$
are small.

\section{Disappearing baryon number}

Our main purpose is to consider configurations with 
non-vanishing baryon number sitting on top of the monopole.
We shall see that the baryon number need not be integer, and
that there are no
static classical solutions with non-zero baryon number.
The strategy to deal with such a situation is standard
\cite{affleck,ambjorn,yanagida}: we shall
impose a constraint ensuring that the baryon number takes on 
a non-vanishing value, find a static classical solution of the 
constrained system and study the behaviour of energy as a
function of the baryon number.

\subsection{Spherically symmetric Ansatz}

We shall seek spherically symmetric solutions of the
constrained system. To write down the Ansatz, we note that the 
action (\ref{action}) is invariant under the spatial reflection,
$x^0 \to x^0$, ${\bf x} \to - {\bf x}$ supplemented by the 
interchange of the left and right $SU(2)$, i.e.,
\begin{eqnarray}
   A_0 &\leftrightarrow& B_0\; , \;\;\; A_i \to -B_i\; ,
  \;\;\; B_i \to - A_i \nonumber \\
   \Phi_A &\to& - \Phi_B\; , \;\;\;  \Phi_B \to - \Phi_A
\nonumber \\
   U &\to& U^{\dagger}
\label{discretesymmetry}
\end{eqnarray}
In the case of static fields with $A_0 = B_0 = 0$,
the most general spherically symmetric Ansatz consistent 
with this discrete  symmetry\footnote{In
our numerical study we used the most general spherically symmetric
Ansatz without imposing the symmetry (\ref{discretesymmetry}). 
We have found, however, that the solutions always have this symmetry.}
 is
\begin{eqnarray}
A_i &=& -{i\over 2}
\left[ \left( \frac{a_1 (r)-1}{r}\right)\varepsilon_{ijk}\sigma_j \hat x_k
+\left( \frac{a_2 (r)}{r} \right) 
(\sigma_i -\hat x_i \hat x \cdot \vec \sigma )
+\left( \frac{a_3 (r)}{r} \right) 
\hat x_i \hat x \cdot \vec \sigma \right] \nonumber \\
B_i &=& -{i\over 2}
\left[ \left( \frac{a_1 (r)-1}{r}\right)\varepsilon_{ijk}\sigma_j \hat x_k
-\left( \frac{a_2 (r)}{r} \right) 
(\sigma_i -\hat x_i \hat x \cdot \vec \sigma )
-\left( \frac{a_3 (r)}{r} \right) 
\hat x_i \hat x \cdot \vec \sigma \right]\:
\nonumber \\
\Phi_A &=& \Phi_B = i v h(r) \hat x \cdot \vec \sigma
\nonumber \\
U&=&\cos f(r) +i\hat x \cdot \vec \sigma \sin f(r)
\label{6m**}
\end{eqnarray}
where $\hat x$ is the unit radius-vector.
With this Ansatz, regularity of the Skyrme field
at the origin requires that $f(0)$ is an integer 
multiple of $\pi$;
without loss of generality we set
\[
    f(0) = \pi
\]
Other conditions, ensuring that the fields are regular at
the origin, are
\[
  a_1 (0) = 1\;, \;\;\;  a_2 (0) = a_3 (0) =0 \;, \;\;\; 
  h(0) = 0
\]

It is convenient to introduce the dimensionless coordinate,
\[
   \rho = gvr
\]
With this coordinate, the distances are measured essentially
in units of the size of the monopole core. The static energy 
functional is then written as follows (neglecting the WZW
term in eq.~(\ref{action}))
\[
  H =  \frac{4\pi v}{g}  \int~d\rho~\cal{E}(\rho)
\]
\begin{eqnarray}
  \cal{E} &=& 
\left[ \left( a_1'+\frac{a_2 a_3}{\rho}\right)^2
+\left( a_2'-\frac{a_1 a_3}{\rho}\right)^2 +\frac{1}{2\rho^2}( a_1^2 
+a_2^2 - 1)^2
\right] \nonumber \\
&+&
2 \left[\rho^2 (h')^2 +2 (a_1^2 +a_2^2) h^2
+\frac{{\lambda}}{4g^2} \rho^2(h^2-1)^2\right]\nonumber \\
&+& \kappa_1 (X^2 + 2Y^2) + \frac{4\kappa_2}{\rho^2} Y^2 (2X^2 + Y^2)
\label{energyrad}
\end{eqnarray}
where the prime denotes the derivative with respect to $\rho$, 
\begin{eqnarray}
    X &=& \rho f' - a_3
   \nonumber \\
    Y &=& a_1 \sin f - a_2 \cos f
\label{defXY}
\end{eqnarray}
and we introduced the dimensionless parameters
\begin{eqnarray}
  \kappa_1 &=& \frac{F_{\pi}^2}{8 v^2}
\nonumber \\
  \kappa_2 &=& \frac{g^2}{64e^2}
\label{kappas}
\end{eqnarray}
It is worth noting that in terms of these parameters, the ratios 
of the sizes and masses of the monopole and Skyrmion,
as given by eqs.~(\ref{12*}) -- (\ref{VV}), are expressed as
\begin{eqnarray}
  \frac{r_{mon}}{r_{Sk}} &=& 0.084 \cdot \sqrt{\frac{\kappa_1}{\kappa_2}}
\label{rtor}\\
  \frac{M_{Sk}}{M_{mon}} &=& 27 \cdot \sqrt{\kappa_1 \kappa_2}
\label{MtoM}
\end{eqnarray}
where we take ${\lambda} = 0.5 g^2$ for definiteness.
These ratios must be small in the limit of point-like monopole.

The energy functional (\ref{energyrad}) is still invariant under
radial gauge transformations which shift $f(\rho)$ and $a_3(\rho)$,
\begin{eqnarray}
    f(\rho) &\to& f(\rho) + \alpha(\rho) \nonumber \\
    a_3(\rho) &\to& a_3(\rho) + \rho \partial_\rho \alpha(\rho)
\label{restgauge}
\end{eqnarray}
and rotate $(a_1,a_2)$. This gauge symmetry is the
remnant of the axial $U(1)$ (we have fixed the vectorial $U(1)$
by imposing the symmetry (\ref{discretesymmetry})). We shall 
fix the remaining gauge freedom shortly.

It follows immediately 
from eqs.~(\ref{energyrad}) and (\ref{defXY})
that in the absence of the monopole, i.e., when
$a_1(\infty) = 1$, $a_2 (\infty) = 0$, the convergence of the
term proportional to $Y^2$ in the energy integral requires that
$f(\infty)$ is an integer multiple of $\pi$. This means that 
the winding number of the Skyrme field is integer, and hence 
is conserved. On the other hand, in the presence of the monopole, 
one has $a_1(\infty) = a_2(\infty) = 0$, and the convergence 
of energy does not impose any constraints on $f(\infty)$.
In the presence of the monopole, the winding number of the Skyrme
 field is arbitary.

For spherically symmetric fields, the gauge-invariant 
baryon number takes the form
\[
   B= \int_0^{\infty}~d\rho~ b(\rho)
\]
\begin{eqnarray}
 b(\rho) &=& \frac{1}{\pi}  \frac{d}{d\rho}
\left[ -f + \frac{1}{2} (a_1^2 - a^2_2) \sin 2f 
- a_1 a_2 \cos 2f \right]
\nonumber \\
&+& \frac{1}{\pi} \left[(a_1a_2' - a_1' a_2) 
- \frac{a_3}{\rho} (a_1^2 + a_2^2 -1) \right]
\label{bnumber}
\end{eqnarray}
Like the winding number of the Skyrme field,
the baryon number too need not be integer.

\subsection{The constrained system}

To consider configurations with non-vanishing baryon number,
it would seem natural to impose a constraint involving the baryon
number, e.g., 
\[
  \int_0^{\infty}~d\rho~ b(\rho) =\mbox{fixed}
\]
However, we have chosen a slightly different constraint which
simplifies the equations. We note first that one of the
Euler--Lagrange equations following from eq.~(\ref{energyrad})
does not contain the second derivatives. This equation is
\begin{equation}
   \rho \frac{\delta H}{\delta a_3} = 0
\label{dhda3}
\end{equation}
In fact, this equation is not completely independent: its 
derivative with respect to $\rho$ is a consequence of the other
Euler--Lagrange equations (this is  analogous to the
time derivative of Gauss' law). If $\delta H/ \delta a_3 $ is zero
at one value of $\rho$, then it is zero everywhere, provided the
second-order Euler--Lagrange equations are satisfied. The idea is
to impose a constraint which would modify the resulting equations
in such a way that the left hand side of eq.~(\ref{dhda3}) be
allowed to be a non-vanishing constant, while the second order
equations remain intact. The suitable form of this constraint is
\begin{equation}
  -\frac{1}{\pi} \int_0^{\infty}~d\rho~\left(f' 
-\frac{a_3}{\rho} \right) = \mbox{fixed}
\label{42+}
\end{equation}
Indeed, upon adding this constraint with the Lagrange multiplier
$\Lambda$ to the Hamiltonian (\ref{energyrad}), one finds that the
only change in the resulting equations is that eq.~(\ref{dhda3}) 
now becomes
\begin{equation}
   \rho \frac{\delta H}{\delta a_3} + \frac{\Lambda}{\pi}= 0
\label{dhda3L}
\end{equation}  
The procedure is then to disregard the latter equation altogether,
solve the original second order Euler--Lagrange equations and
find $\Lambda$ from eq.~(\ref{dhda3L}) (the last step is actually
unnecessary).

We note that the constraint (\ref{42+}) is invariant under gauge
transformations (\ref{restgauge}). 
In fact, this constraint can be written
in fully gauge-invariant form, since
\[
  \left(f' - \frac{a_3}{\rho} \right) \propto \hat{x}_i 
\mbox{Tr} \left( \frac{\Phi_A}{|\Phi_A|} D_i U U^{\dagger} \right)
\]
Note also that the expression on the left hand side of 
eq.~(\ref{42+}) coincides with the expression for the baryon 
number, eq.~(\ref{bnumber}), except for terms in the latter
containing $a_1$ and
$a_2$. This indicates that the constraint (\ref{42+}) should
give rise to non-vanishing baryon number (indeed, we shall see
below that the integral (\ref{42+})
approximates the baryon number very well).

The fact that one no longer has to consider the variations of
$a_3$ simplifies the analysis considerably. We can now impose
the gauge condition
\[
   a_3 = 0
\]
directly in the Hamiltonian. Thus, the problem reduces to solving 
the Euler--Lagrange equations following from the functional
 \[
  \tilde{H} \equiv H(a_3=0) 
=  \frac{4\pi v}{g}  \int~d\rho~\tilde{\cal{E}}(\rho)
\]
\begin{eqnarray}
  \tilde{\cal{E}} &=& 
\left[ ( a_1')^2
+( a_2')^2 +\frac{1}{2\rho^2}( a_1^2 
+a_2^2 - 1)^2
\right] \nonumber \\
&+&
2 \left[\rho^2 (h')^2 +2 (a_1^2 +a_2^2) h^2
+\frac{{\lambda}}{4g^2} \rho^2(h^2-1)^2\right]\nonumber \\
&+& \kappa_1 (\rho^2 (f')^2 + 2Y^2) 
+ \frac{4\kappa_2}{\rho^2} Y^2 (2\rho^2 (f')^2 + Y^2)
\label{energyrad1}
\end{eqnarray}  
where $Y$ is defined in eq.~(\ref{defXY}). All these 
equations are of  second order; their form is not very
illuminating, and we do not reproduce all of them 
here.
%\footnote{\large{\bf End of the part read by everybody.
%First equation below was checked by Yves.}}
We only write down the equation which is obtained by varying 
$\tilde{H}$ with respect to $f$,
\begin{equation}
  \frac{\partial}{\partial \rho}\left[ (\kappa_1 \rho^2 +
8 \kappa_2 Y^2) f'\right] = 
2 \left[\kappa_1 + 4\kappa_2 (f')^2 + 
4 \frac{\kappa_2}{\rho^2}Y^2 \right] (a_1 \cos f + a_2 \sin f)Y
\label{a*}
\end{equation}
It is straightforward to see that $a_1$ and $a_2$ decay
exponentially at large $\rho$,
\[
  a_1, a_2 \propto C \mbox{e}^{-2\rho} + 
C' \mbox{e}^{ -\sqrt{4 + 2 \kappa_1} \rho}
\]
Hence, away from the monopole core, equation (\ref{a*}) becomes
\[
   (\rho^2 f')' =0
\]
and its solution is
\begin{equation}
   f(\rho) = f_{\infty} + \frac{c_f}{\rho}
\label{a**}
\end{equation}
where $f_{\infty}$ and $c_f$ are constants. In fact, only the 
asymptotic value $f_{\infty}$ is a free parameter (for given
$\kappa_1$ and $\kappa_2$), as the constant $c_f$ is a function
of $f_{\infty}$ for the soultions of the complete system.

Thus, the system of Euler--Lagrange equations, corresponding to the 
functional (\ref{energyrad1}), 
admits a one-parameter family of solutions
parametrized by the asymptotic value $f_{\infty}$. The baryon number 
and the energy for the solutions are determined by $f_{\infty}$.

\subsection{Numerical results}

We have obtained these families of solutions numerically for various
values of $\kappa_1$ and $\kappa_2$ and computed their energies and
baryon numbers. We shall present our results for a fixed value of
the Higgs self-coupling,
\[
{\lambda}/g^2 = 0.5
\]
To a good accuracy the baryon number is a linear function
of $f_{\infty}$: according to eq.(\ref{bnumber}), in the gauge
$a_3=0$ we have
\begin{equation}
   B = \frac{\pi - f_{\infty}}{\pi} + 
\frac{1}{\pi} \int_0^{\infty}~d\rho~(a_1 a_2^{\prime} -
a_1^{\prime} a_2)
\label{bv*}
\end{equation}
The last integral here is small; we shall discuss this property later on.

Our first result is that the energy is always a monotonic function of
the baryon number. This is shown in Fig. 1 for three sets of parameters 
$(\kappa_1, \kappa_2)$. This means that there are no static solutions
with non-vanishing baryon number on top of the monopole.

We note that the third set of parameters, $\kappa_1 = 10^{-4}$,
$\kappa_2 = 0.1$ corresponds to a very small size of the monopole
core as compared to the Skyrmion size, 
$r_{mon} / r_{Sk} = 2.7 \cdot 10^{-3}$.
It is natural to expect that in this case, the regime of 
``point-like'' monopole is realized. Let us discuss this
in some detail, with the of purpose of estimating the range of parameters 
at which this regime sets in.

The situation of interest is when the monopole is small and heavy,
\[
    \frac{r_{mon}}{r_{Sk}} \ll 1\; , \;\;\;\;
    \frac{M_{mon}}{M_{Sk}} \gg 1
\]
This situation occurs at small $\kappa_1 / \kappa_2$ and
$\kappa_1 \cdot \kappa_2$, see eqs. (\ref{rtor}), (\ref{MtoM}).
In this case the monopole core is unaffected by the Skyrme field,
i.e., the profiles of $a_1(\rho)$ and $h (\rho)$ coincide with the
undistorted monopole profiles (and hence are independent of the baryon
number), whereas $a_2(\rho)$ is small. The first two terms in 
eq.~(\ref{energyrad1}) are then irrelevant, and the properties of the
radial Skyrme field $f$ are determined by the Hamiltonian
\begin{eqnarray}
  (E - M_{mon})[f] &=& 4\pi \frac{F_{\pi}}{\sqrt{512}e}
\int~d\rho~ (\tilde{\cal{E}} - \tilde{\cal{E}}_{mon})(\rho)
\nonumber \\
\tilde{\cal{E}} - \tilde{\cal{E}}_{mon} &=&
\sqrt{\frac{\kappa_1}{\kappa_2}} \left[ \rho^2 (f')^2 + 2 Y^2\right]
+ 4 \sqrt{\frac{\kappa_2}{\kappa_1}} \frac{Y^2}{\rho^2}
\left[ 2 \rho^2 (f')^2 + Y^2\right]
\label{f+}
\end{eqnarray}
where we made use of eq.~(\ref{kappas}). The fields $a_1$ and $h$
entering this expression are spectators (they correspond
to the undistorted monopole). 
This is what we mean by the regime of ``point-like'' 
monopole. In fact, the monopole is in some sense
never point-like, 
as its core affects the asymptotics of the radial Skyrme field (this
feature
we shall discuss later on), so it is more appropriate to speak of
a {\it spectator monopole}.

There are several properties of the {\it spectator monopole} regime
that may be used to estimate the range of parameters
at which it is actually realised: 

(i) the profiles of $a_1$ and $h$
must coincide with the undistorted monopole profiles;

(ii) the profile $f(\rho)$ should depend on the {\it ratio} 
$\kappa_1/\kappa_2$, and not on the overall magnitude of $\kappa_{1,2}$;
in particular, for fixed $f_{\infty}$, the second asymptotic coefficient 
$c_f$ in eq.~(\ref{a**}) should depend on $\kappa_1 / \kappa_2$ only
(this follows from eq.~(\ref{f+}));

(iii) the baryon number should be equal to 
$(\pi - f_{\infty})/ \pi$ (this follows from eq.~(\ref{bv*}) and
$a_2=0$);

(iv) The energy, referenced from the monopole mass and expressed in
units of the Skyrmion mass, must depend only on the baryon number
and {\it ratio} $\kappa_1/\kappa_2$,
\[
   \frac{E - M_{mon}}{M_{Sk}} 
= \Delta E \left(B, \frac{\kappa_1}{\kappa_2}\right)
\]
(this again follows from eq.~(\ref{f+})). 

The properties (ii) -- (iv) are illustrated in Figs.~2~--~4.

The property (i) is particularly useful for establishing at what 
values of $\kappa_1$ (i.e., $F_{\pi}/{v}$) the spectator monopole 
regime sets in for given $\kappa_2$ (i.e., for given $g^2 / e^2$).
We present the corresponding plots in 
Fig.~5.

Figures~2~--~5 show that the spectator monopole regime sets in
at fairly large $\kappa_1$ and $\kappa_2$, as large as 
$\kappa_1 = 0.01$, $\kappa_2 = 0.1$. These values of parameters
correspond to $r_{mon} \sim 0.03 r_{Sk}$ and 
$M_{Sk} \sim M_{mon}$. A ``point-like'' monopole need not be
really very small and very heavy.

\subsection{Radial Skyrme field}

Let us now discuss how the asymptotics of the radial Skyrme 
function, namely, the constant $c_f$ 
 at given $f_{\infty}$, 
depends on the properties of the monopole
core in the spectator monopole regime. At small 
$\kappa_1 / \kappa_2$, the behaviour (\ref{a**}) terminates at 
$\rho > 1$ where the monopole function $a_1$ is already 
exponentially decaying (i.e., somewhat outside the core).
Indeed, the term
$8 \kappa_2 Y^2$ on the left hand side of eq.~(\ref{a*})
becomes comparable to $\kappa_1 \rho^2$ at
\begin{equation}
    a_1 \sin f \sim \sqrt{\frac{\kappa_1}{\kappa_2}} \rho
\label{m*}
\end{equation}
This implies that the regime (\ref{a**}) terminates when $a_1$ is 
small, and $\rho \sim |\ln a_1|/2 $ is larger than 1. For a crude
estimate we take into account this effect only, set $\sin f \sim 1$
on the left hand side and $\rho \sim 1$ on the right hand side of
eq.~(\ref{m*}) and obtain that the asymptotic regime
(\ref{a**}) terminates at
\begin{equation}
   \rho \sim \rho_c = \frac{1}{4} 
| \ln \left(\frac{\kappa_1}{\kappa_2}\right)|
\label{13m*}
\end{equation}
At this value of $\rho$, the function $f(\rho)$ should be reasonably 
close to its value at $\rho=0$, i.e., $f(\rho_c) \sim \pi$. Hence
\[
   \frac{c_f}{\rho_c} \sim (\pi - f_{\infty})
\]
that is
\[
   c_f \sim \frac{1}{4} (\pi - f_{\infty}) 
| \ln \left(\frac{\kappa_1}{\kappa_2}\right)|
\]
up to terms varying slower than logarithmically.
This shows that $c_f$ depends on the ratio $\kappa_1/\kappa_2$
logarithmically. We confirmed
this expectation by numerical calculations, though 
instead of the coefficient $\pi/4$ at $f_{\infty}=0$, we obtained that
in the spectator monopole regime
\[
   c_f = 0.61 \cdot  | \ln \left(\frac{\kappa_1}{\kappa_2}\right)| 
+ \mbox{const} \; , \;\; \; \; f_{\infty} = 0
\]
Thus, in terms of the physical coordinate $r$, the radial Skyrme function
outside the core behaves as
\begin{equation}
   f(r) = f_{\infty} + \mbox{const}\cdot 
\frac{r_{mon} |\ln (r_{mon}/r_{Sk})|}{r}
\label{14m*}
\end{equation}
Everywhere in space, except for a region close to the monopole core,
the least energy configuration with given baryon number is 
the $r$-independent hedgehog field,
$f = \cos f_{\infty} + i \hat{x} \cdot \vec \sigma \sin f_{\infty}$,
while $f(r)$ changes from $\pi$  to $f_{\infty}$ 
near $\rho \sim \rho_c$.

A crude estimate for the energy of configurations with, say,
$B=1$ is obtained by integrating the energy density,
$(F_{\pi}^2/8)\cdot r^2 (f')^2$, over the region 
$r \ge r_c \equiv gv \rho_c$. One finds, using the asymptotics
(\ref{a**})
\[
  (E - M_{mon}) \sim \frac{F_{\pi}^2}{gv} \frac{c_f^2}{\rho_c}
\]
i.e., up to logarithms,
\begin{equation}
 (E - M_{mon}) \sim \frac{F_{\pi}^2}{gv}
\label{+*}
\end{equation}
Hence, the Skyrmion looses most of its mass when approaching the
monopole. This is illustrated in
Fig. 6.

We note in passing that if the baryon number were
conserved, there would exist Skyrmion--monopole bound states
with binding energy almost equal to the mass of the Skyrmion.
This is in accord with Refs.~\cite{callanwitten,tigran1,tigran2}.

\subsection{Effect of Wess--Zumino--Witten term}

Let us now discuss the Wess--Zumino--Witten
 term in the action, the
last term in eq.~(\ref{action}). In our context, its 
main effect is to generate charge densities induced by the
Skyrme and gauge fields \cite{Goldstone:1981kk}. A suitable 
gauge for calculating this effect is $U=1$. In this gauge,
the WZW term is \cite{wittenwz} (we write the terms 
which do not vanish in the $SU(2)$ case)
\[
\Gamma_{WZW} =
\int~d^4x~ \epsilon^{\mu\nu\lambda\rho} Z_{\mu\nu\lambda\rho} 
\]
\[ 
  Z_{\mu\nu\lambda\rho} = - \frac{n}{48\pi^2} 
\mbox{Tr} \left[ A_\mu A_\nu \partial_{\lambda} B_{\rho} 
 + B_\mu B_\nu \partial_{\lambda} A_{\rho}
+ \frac{1}{2}( A_\mu A_\nu B_\lambda B_\rho
   +  B_\mu A_\nu B_\lambda A_\rho ) \right]
\]
where $n$ is an integer (number of colours in QCD).
A straigtforward calculation gives for general 
spherically symmetric fields
\[
\int~d^3x~ \epsilon^{\mu\nu\lambda\rho} Z_{\mu\nu\lambda\rho}
= \frac{n}{48\pi^2} \int~4\pi r^2dr~ Z
\]
\[
Z=
\frac{\tilde{a}_0}{r^2} \left[ \frac{\tilde{b}_3}{r}
(\tilde{b}_1 \tilde{a}_2 - \tilde{a}_1 \tilde{b}_2)
 - \tilde{b}_{2}'(\tilde{a}_2 - \tilde{b}_2) 
- \tilde{b}_{1}' (\tilde{a}_1 - \tilde{b}_1) \right] + 
(\tilde{a} \leftrightarrow \tilde{b})
\]
where $\tilde{a}_i$, $\tilde{b}_i$ are the radial fields 
in the gauge $U=1$ (in writing this formula we did not
assume the symmetry (\ref{discretesymmetry}); the field
$B_i$ has the same form as $A_i$ in (\ref{6m**}) 
but with $b_i$ substituted for 
$a_i$), and 
\[
\tilde{a}_0 = i \mbox{Tr} (\hat{x}\cdot \vec \sigma A_0) \; , \;\;\;
\tilde{b}_0 = i \mbox{Tr} (\hat{x}\cdot \vec \sigma B_0)
\]
%Returning to the original gauge $a_3 = b_3 = 0$, and
Making use of the symmetry (\ref{discretesymmetry}), we find
\begin{equation}
  Z = \frac{2(\tilde{a}_0 + \tilde{b}_0)}{r^2}
\left( \tilde{a}_2 \tilde{a}_{2}' - 
\frac{\tilde{a}_1 \tilde{a}_2 \tilde{a}_3}{r} \right)
\label{N2*}
\end{equation}
Hence, the WZW term induces (vectorial) electric
charge density in the vicinity of the monopole.

We have to check that the Coulomb self-interaction
due to this effect does not spoil our previous analysis.
Let us estimate the Coulomb energy, making use of the
spectator monopole approximation. 
Returning to the original gauge $a_3 = b_3 = 0$, and
setting $a_2 = b_2 = 0$, we obtain
%In this approximation,
%eq.~(\ref{N2*}) becomes
\[
  Z = \frac{2(a_0 + b_0)}{r^2} a_1 a_{1}' \sin^2 f
\]
The electric charge density is induced in the region
where $f$ is substantially different from $\pi$.
This occurs near and above 
$\rho \sim \rho_c$, where $a_1$ is small.
Making use of eq.~(\ref{m*}) we find that the total 
induced charge is, up to logarithms,
\[
  Q \sim [a_1(\rho_c) \sin f (\rho_c)]^2  \sim
\frac{\kappa_1}{\kappa_2} \sim \frac{e^2 F_{\pi}^2}{g^2v^2}
\]
Since $r_c \sim r_{mon} \sim (gv)^{-1}$, up to logarithms,
we obtain an estimate for the Coulomb energy
\[
  E_{Coul} \sim \frac{g^2 Q^2}{r_c} 
\sim \frac{F_{\pi}^2}{gv} \left( \frac{e^2 F_{\pi}}{v} \right)^2
\]
which is very small compared to eq.~(\ref{+*}).

Hence, in our model, the Skyrmion on top of the monopole
is (nearly) neutral, and the Coulomb energy is negligible.
The WZW term in the action may be safely ignored.

\section{Discussion}

For fixed parameters of the model, $\kappa_1$ and $\kappa_2$,
we have constructed the family of constrained solutions along which
the baryon number monotonically decreases to zero and energy 
monotonically decreases to the monopole mass. These solutions, 
however, have different behaviours of the Skyrme field at spatial
infinity, i.e., different values of the radial Skyrme function at
$r=\infty$. One may doubt that the system can actualy travel along this
path in configuration space: in the gauge $A_0 = B_0 = 0$,
the kinetic energy contains a term
\[
   \frac{\pi F_{\pi}}{2} \int~r^2dr~ \dot{f}^2
\]
which diverges if $f_{\infty}$ changes in time. Also, the baryonic 
current (\ref{8}) has spatial components whose asymptotics at large 
$r$  (again in the gauge $A_0 = B_0 =0$) are
\begin{equation}
   j_i = \frac{\hat{x}_i}{4\pi^2 r^2} \dot{f}
\label{q*}
\end{equation}
For $\dot{f}_{\infty} \neq 0$, there is the baryonic flux at
$r \to \infty$,
\[
   F = \int~j_i~d\Sigma^i = \frac{1}{\pi} \dot{f_{\infty}}
\]
This may seem to indicate 
that the bayon number leaks to spatial infinity.

Certainly, an evolution with infinite kinetic energy is impossible,
so in the gauge $A_0 = B_0 = 0$ the asymptotic value $f_{\infty}$
cannot change in time. To see how, in this gauge, the system can travel
along the path with varying baryon number (say, leading from $B=1$
to $B=0$), one has first to transform each constrained solution
of our family to a gauge in which $f(r)$ rapidly tends to zero
as $r \to \infty$. According to eqs. (\ref{restgauge}), (\ref{a**}),
each configuration of the new family will have non-vanishing $a_3$,
so that at large $\rho$ 
\begin{equation}
   a_3 = - \frac{c_f}{\rho}
\label{r*}
\end{equation}
Because of gauge invariance, the baryon numbers and energies of 
old and new configurations will be the same.

Travelling along the new path in the configuration space does not
cost infinite kinetic energy. Indeed, in the gauge $A_0=B_0=0$,
the electric field at large $r$ behaves as
\[
   F_{0i} \propto \frac{\dot{a}_3}{r} \propto \frac{\dot{c}_f}{r^2}
\]
and the integral of $F_{0i}^2$ converges. Also, the baryonic current 
(\ref{q*}) vanishes as $r \to \infty$, so the baryon number does not 
leak to spatial infinity. It merely disappears.

\bigskip

\noindent
{\bf Acknowledgements}

This work was carried out in the framework of projects SC/2000/020 and
IC/2000/021 of Enterprise-Ireland.
The work of V.R. has been supported in part by the Council for
Presidential Grants and State Support of Leading Scientific Schools,
grant 00-15-96626, RFBR grant 99-01-18410 and CRDF grant (award RP1-2103).

\newpage

{\large {\bf Figure captions}}

\vspace{5mm}

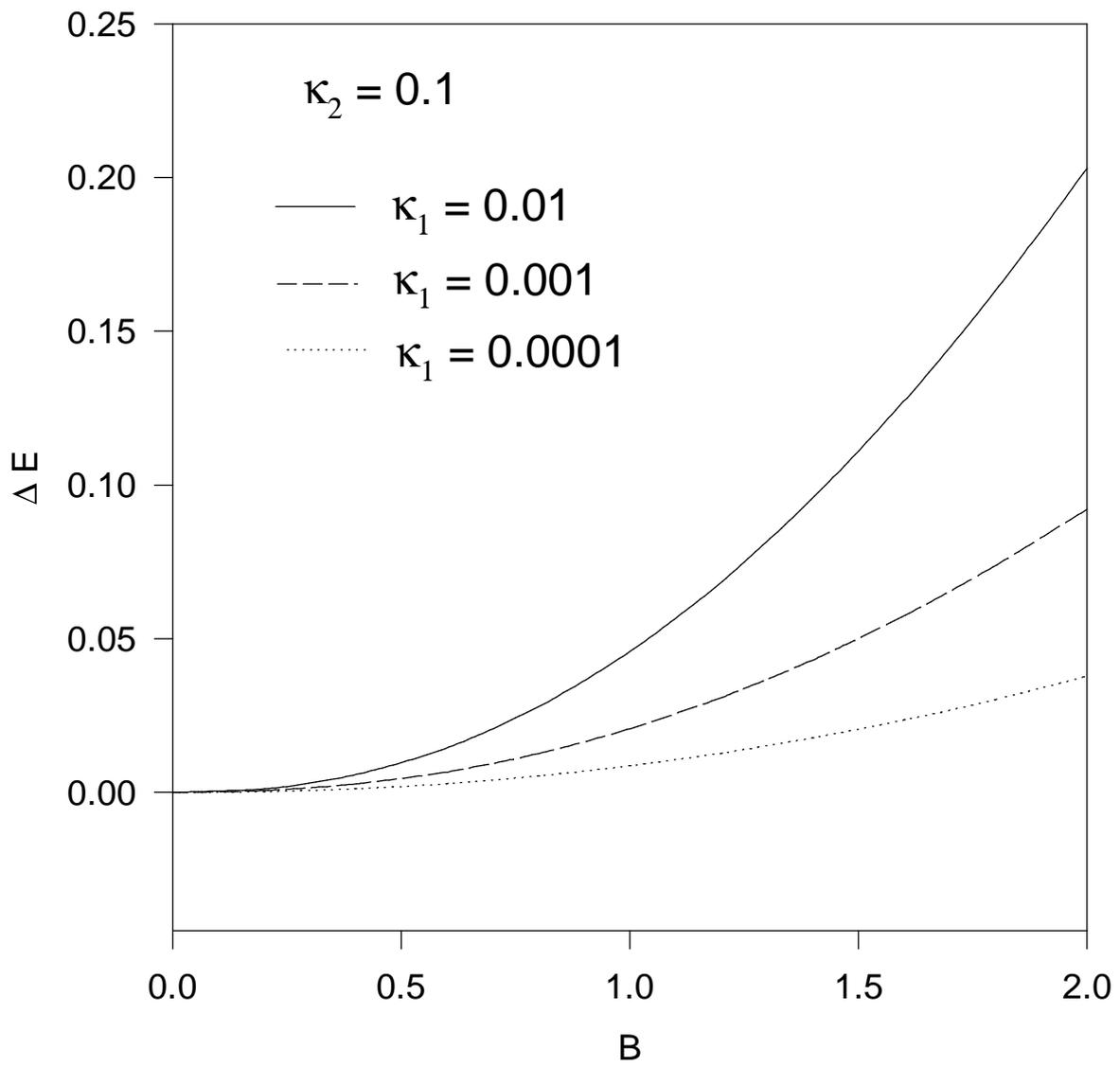
{\bf Fig 1.} Energy (referenced from the monopole mass
and expressed in units of the Skyrmion mass) versus baryon
number for $\kappa_2 = 0.1$, and $\kappa_1 = 0.01$ (solid line)
$\kappa_1 = 0.001$, (dashed line), $\kappa_1 = 0.0001$ 
(dotted line).

\vspace{5mm}

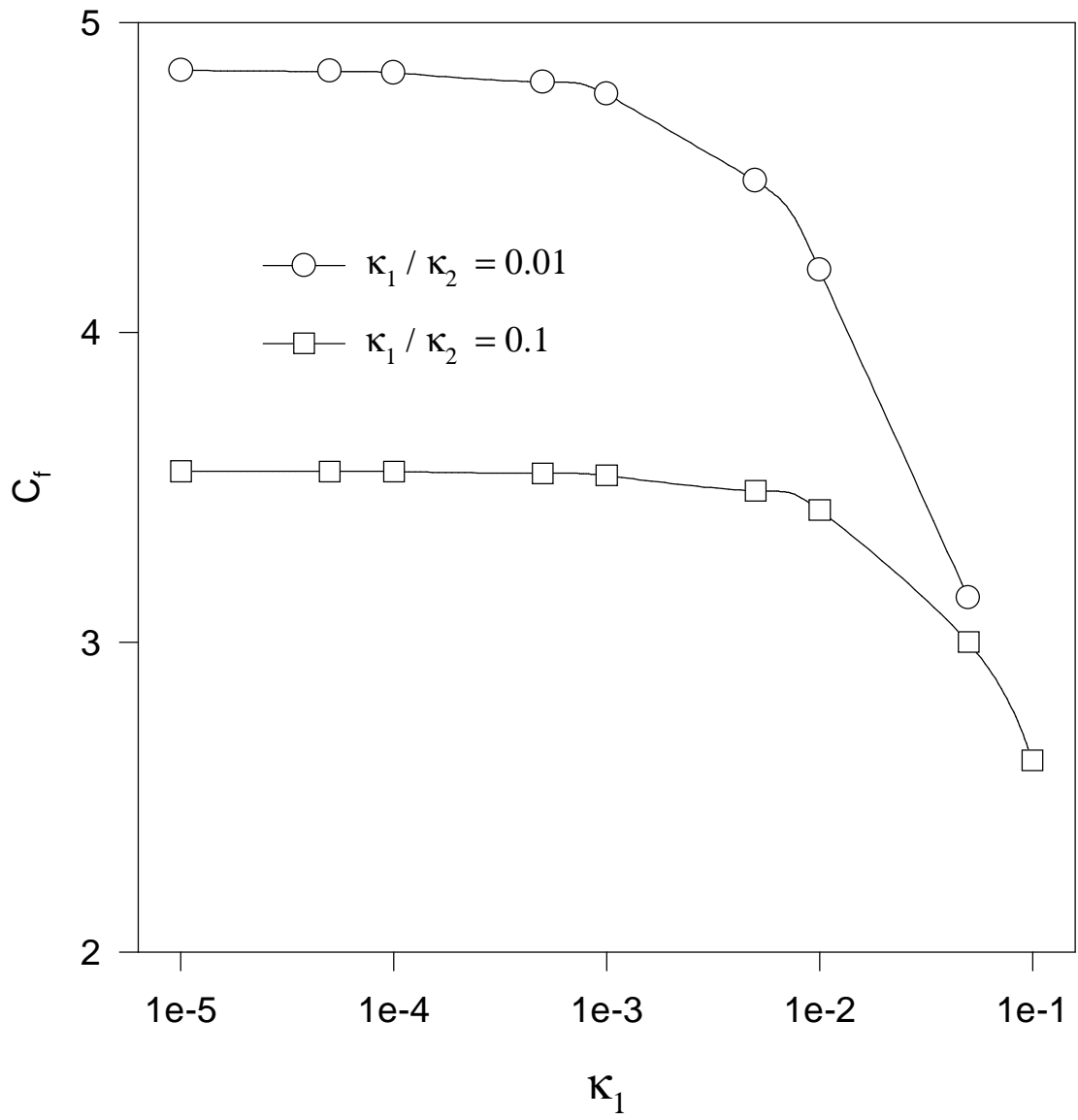
{\bf Figure 2.} The asymptotic coefficient
$c_f$ (see eq.~(\ref{a**}))
at $f_{\infty}=0$ as a function of $\kappa_1$ for fixed ratio
$\kappa_1 / \kappa_2 = 0.1$ (squares) and
$\kappa_1 / \kappa_2 = 0.01$ (bullets).

\vspace{5mm}

{\bf Figure 3.}
The baryon number 
as a function of 
$\kappa_1$ for two values of the ratio
$\kappa_1 / \kappa_2$:

(a) $(B-1)$ at  $f_{\infty} = 0$;

(b) $(B- 0.5)$ at  $f_{\infty} = \pi /2$.

\vspace{5mm}

{\bf Figure 4. }
Energy as a function of baryon number at
different $\kappa_1$ and fixed ratio (a) $\kappa_1 / \kappa_2 = 0.1$;
(b) $\kappa_1 / \kappa_2 = 0.01$. 
The curves for $\kappa_1 \leq 0.0001$
are indistinguishable.

\vspace{5mm}

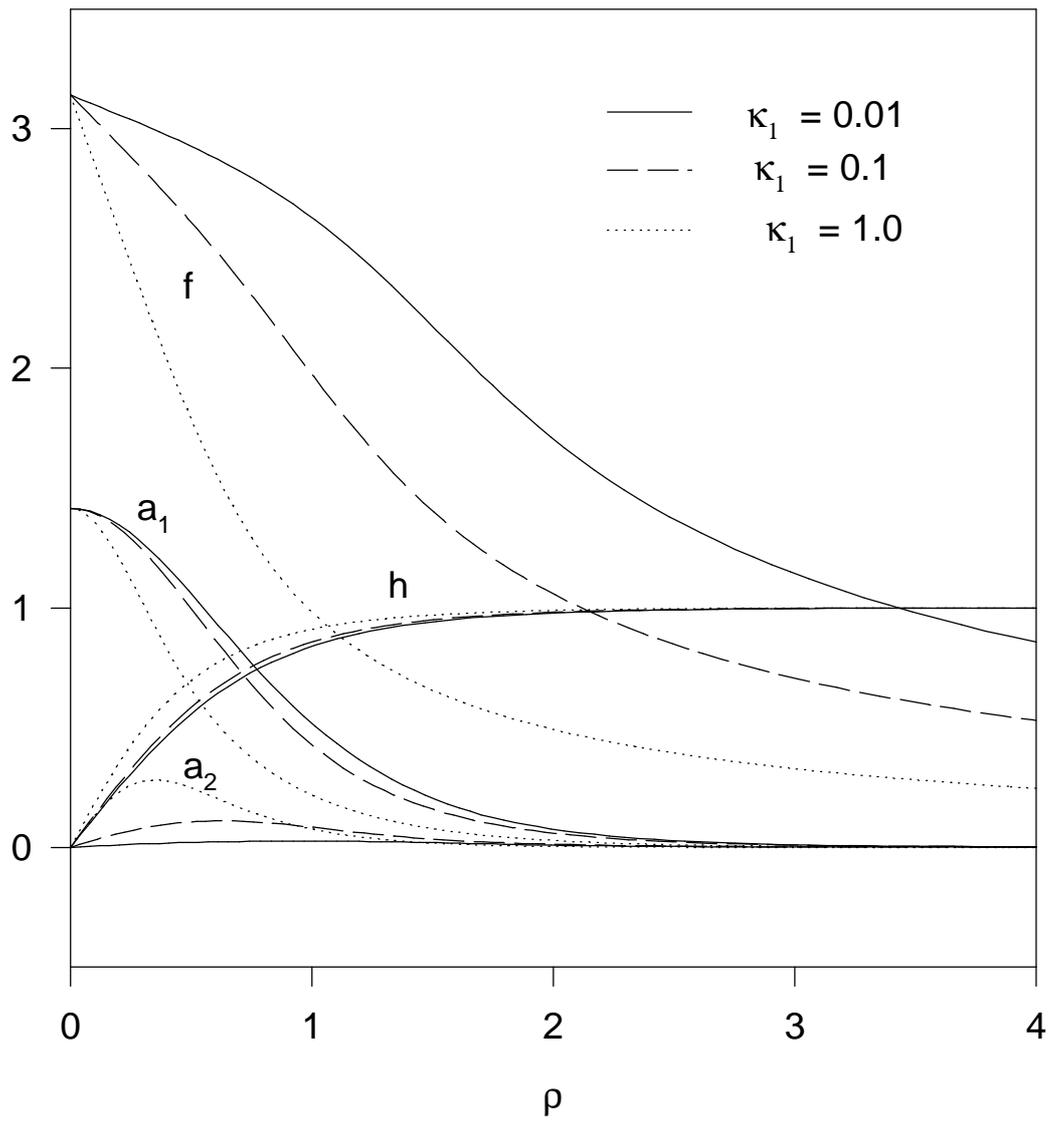
{\bf Figure 5.} Profiles of the functions
$a_1$, $a_2$, $h$ and $f$ for 
$\kappa_2 = 1$ and $\kappa_1 = 1$ (dotted line),
$\kappa_1= 0.1$ (dashed line), $\kappa_1=0.01$ (solid line).
For $\kappa_1 \leq 0.01$ the profiles of $a_1$, $a_2$ and
 $h$ are indistinguishable and
coincide with the undistorted monopole profiles.

\vspace{5mm}

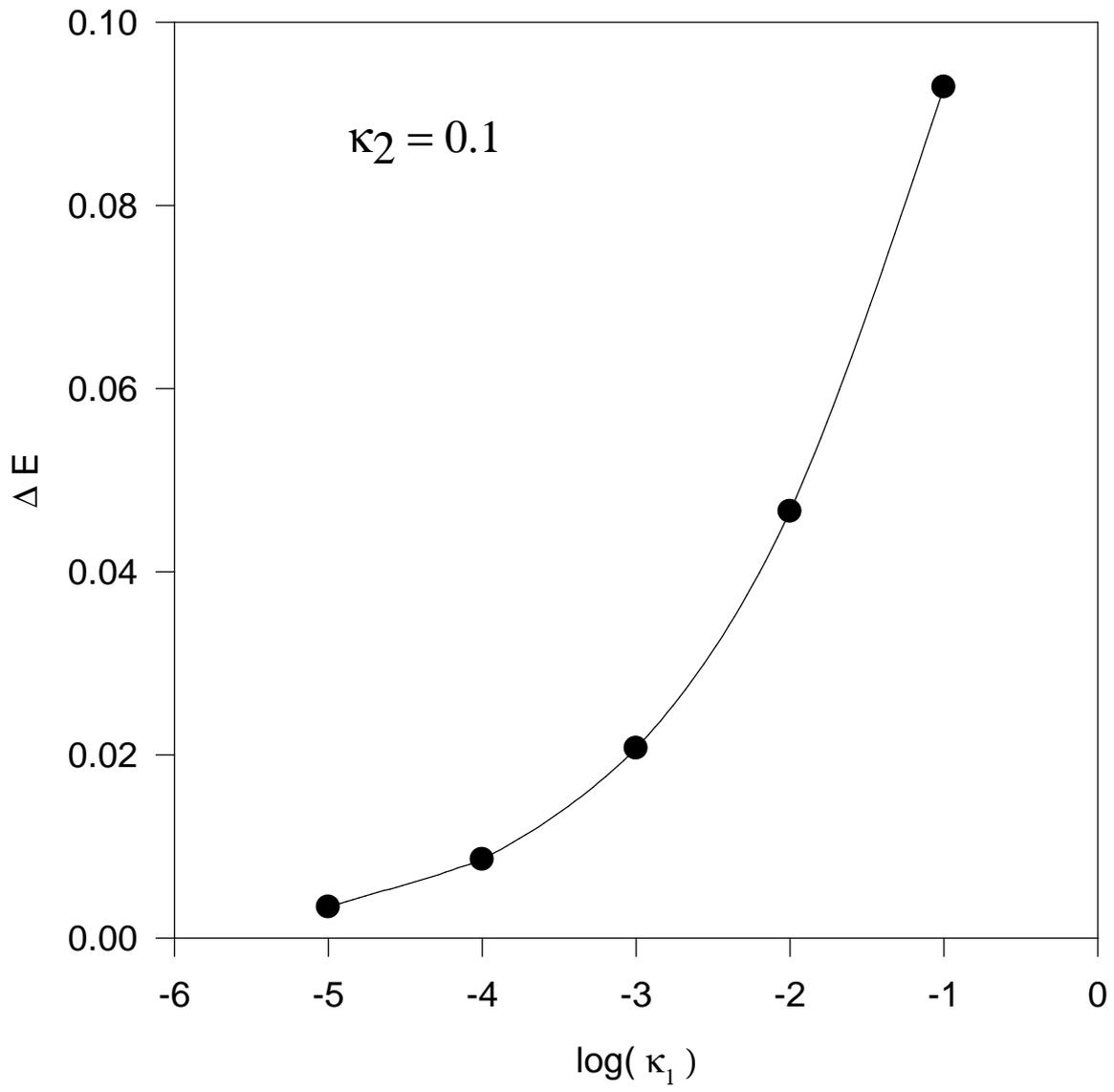
{\bf Figure 6.} 
 Energy (referenced from the monopole mass
and expressed in units of the Skyrmion mass)
for the configurations with baryon number $B=1$
as a function of $\kappa_1$ for a fixed value 
of $\kappa_2$.

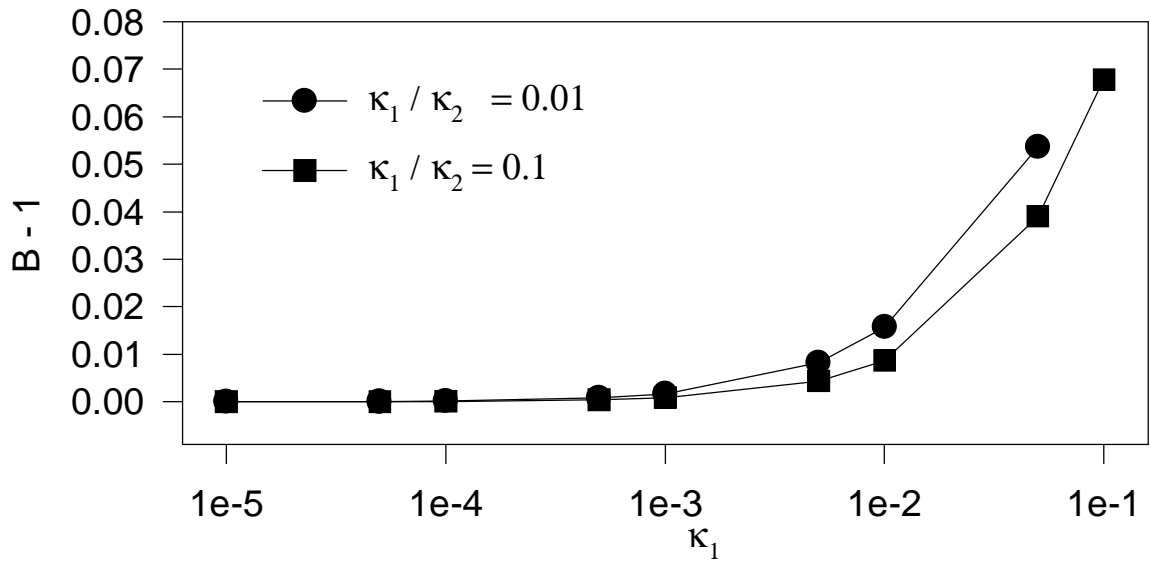
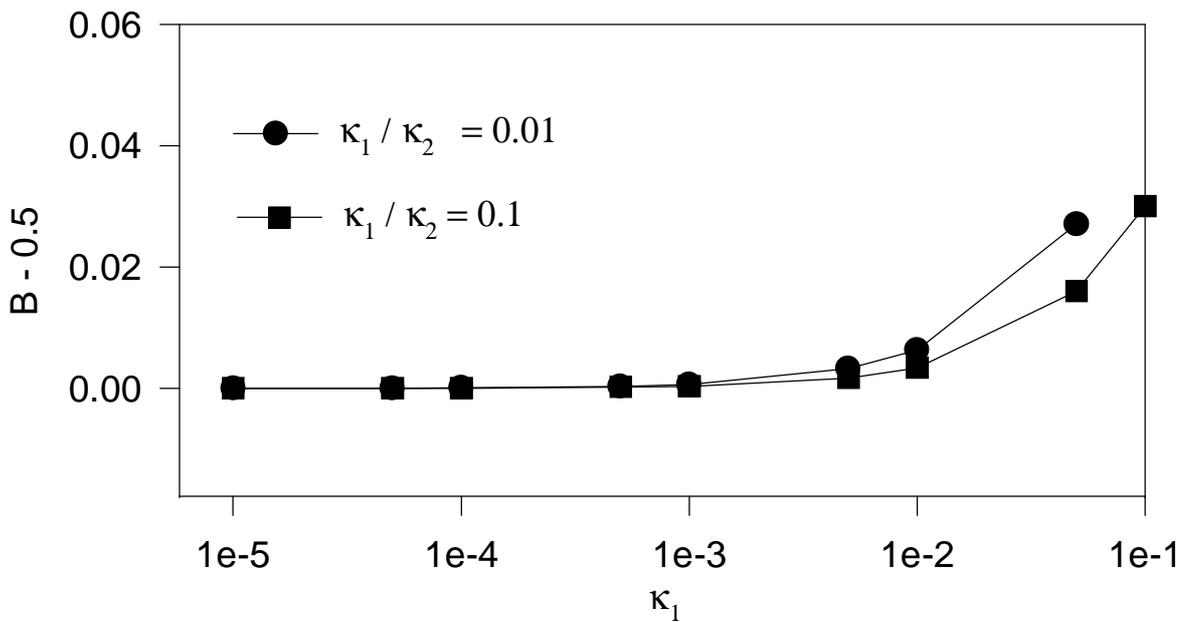
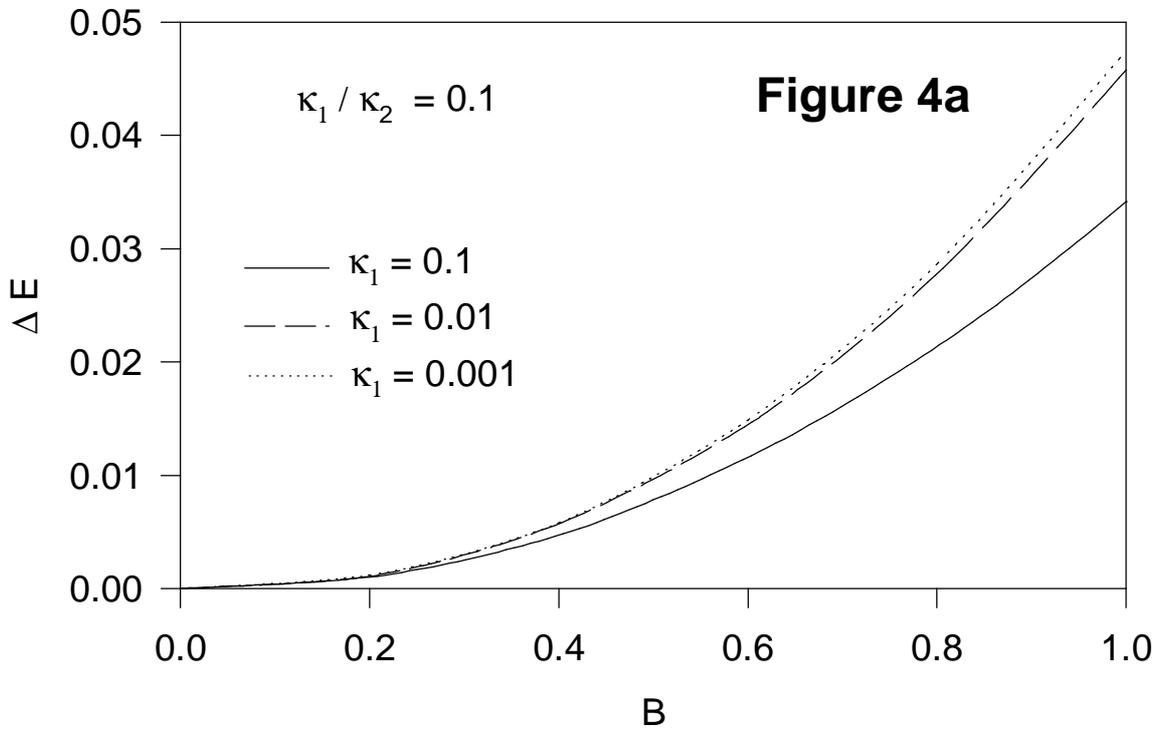
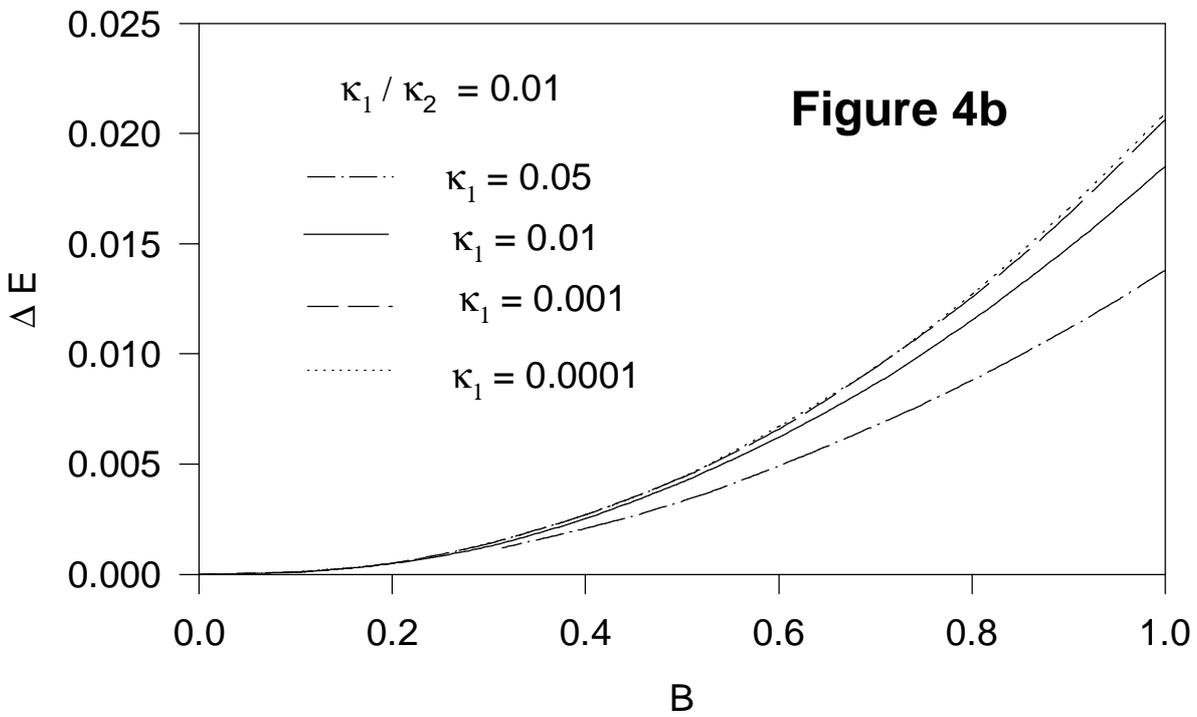
\end{document}